# Analysis of defect formation in semiconductor cryogenic bolometric detectors created by heavy dark matter

Ionel Lazanu [a], Magdalena Lidia Ciurea [b], Sorina Lazanu [b,1]

[a] University of Bucharest, Faculty of Physics, POBox MG-11, Bucharest-Măgurele, Romania
[b] National Institute of Materials Physics, POBox MG-7, Bucharest-Măgurele, Romania

**Abstract**
The cryogenic detectors in the form of bolometers are presently used for different applications, in particular for very rare or hypothetical events associated with new forms of matter, specifically related to searches for Dark Matter. In the detection of particles with a semiconductor as target and detector, usually two signals are measured: ionization and heat. The amplification of the thermal signal is obtained with the prescriptions from the Luke–Neganov effect. The energy deposited in the semiconductor lattice as stable defects in the form of Frenkel pairs at cryogenic temperatures, following the interaction of a dark matter particle, is evaluated and consequences for measured quantities are discussed. This contribution is included in the energy balance of the Luke effect. Applying the present model to germanium and silicon, we found that for the same incident weakly interacting massive particle the energy deposited in defects in germanium is about twice the value for silicon.

## 1. Introduction

In the last decades, great developments in low temperature detectors in the form of bolometers, in the technologies of semiconductors, superconductors or scintillator crystals were obtained. These cryogenic detectors are able to detect radiations and particles with a threshold in the range of eV.

If the pioneering idea of the bolometric detectors goes back to 1935, year when Simon suggested an "Application of Low Temperature Calorimetry to Radioactive Measurements" [1], the modern applications started after the '70. Nowadays, there are a lot of reviews in this thematic; see for example those of Gaitskell [2] and Sarazin [3]. Bolometric detectors are used in different applications in experimental physics, e.g. in searches for neutrinoless double beta decay and neutrino mass - for example the experiments CUORE & Cuoricino, for total energy measurements of free electron lasers [4] to measure the cosmological microwave background [5] constituents of the dark matter, *etc*.

---

[1] Corresponding author: Tel: +40213690185; Fax: +40213690177
Email: lazanu@infim.ro

There is clear evidence that a large part of the dark matter in the Universe is non-baryonic, non-luminous and non-relativistic and the search for it has become a very active research area in the last decades. Hypothetical Weakly Interacting Massive Particles (WIMPs) are proposed as possible particle candidates that satisfy all of the above criteria. Thus, their direct detection using the experimental information of low-energy nuclear recoils originating from WIMPs interactions is one of the detection methods usually used in bolometric detectors.

If in the first generation of these experiments only the heat deposited in detectors as phonons was used in the detection, in more recent experiments phonons and ionization (or light from scintillation signals) are measured simultaneously, trying to discriminate both between electron - nucleon/nuclei recoils and also between different sources of the phenomena: ordinary matter or constituents of the dark matter. As detector materials, silicon and germanium or scintillator crystals ($Al_2O_3$:Ti or $CaWO_4$, $CaMoO_4$ etc.) are used.

One of the effects produced by the slowing down of particles in crystalline semiconductors is defect production, which is a phenomenon present at all temperatures. Defect formation after electron and gamma irradiation at temperatures around and lower than liquid He was studied in InP, Si, Ge, and SiC since 1995 [6 – 8].

In this paper we discuss the effects introduced in the energy balance by the formation of long time stable defects in materials for bolometers and possible consequences for the identification of the particles. In the next section, general aspects related to defect formation in the process of slowing down of selfrecoils in silicon and germanium are reviewed, with emphasis on the existing experimental data related to defect formation following cryogenic irradiation. The energy stored in Frenkel pairs is calculated, and the formulae relating it to the measured quantities in heat and ionization detectors are derived. Concrete applications related to direct WIMPs searches with these detectors are discussed, underlying the influence of the energy stored in defects.

## 2. Defects at cryogenic temperatures and energetic aspects

2.1 Energy balance in heat and ionization cryogenic detectors

After the primary interaction of an incoming particle in the semiconductor, a selfrecoil of energy *E* is left. It loses energy in both electronic and nuclear collisions.

Let $v(E)$ be the energy deposited in the semiconductor in the form of atomic collisions, and $\eta(E)$ the total energy given to the electronic system, both calculated using Lindhard's theory [9]. Part of the energy $v(E)$ is stored in lattice defects ($E_D$), the other part being given to the lattice in the form of excitations (phonons).

If the energy of the recoil is lower than the threshold energy for displacements, this energy is transferred directly to the phonon system. On the other hand, if the energy imparted by the projectile in the primary interaction is high enough, the recoil creates a displacement cascade, composed of equal numbers of vacancies and interstitials, which could also be in the form of Frenkel pairs (FP). Mobile vacancies and interstitials could annihilate.

The number of FPs produced by a recoil depends also on the threshold energy for displacements in the lattice, which in its turn contains, besides the formation energy of a FP, a quantity which goes to the lattice, consisting mostly in a bond-bending component [10], due to the fact that defect formation is a complex multi-body collision process (a small collision cascade) where the atom that receives energy can also bounce back, or kick another atom back to its lattice site.

Defect production following low temperature (~ 4K) electron and gamma irradiation was mainly studied by X ray diffraction methods, using measurements of the change of the lattice parameter and the diffuse scattering of X rays close to different Bragg reflections [11] and by positron annihilation spectroscopy [12]. This way, it was demonstrated that FPs are produced by irradiation at 4 K, and are frozen in both in Si [7, 13, 14] and in Ge [15], at least up to 10 K.

It was shown experimentally that in Si the introduction rates of FPs are independent on the doping type and level and on the growth technique. At temperatures higher than 10 K, they dissociate and/or recombine in more annealing stages, probably related to the distance between the constituents (vacancy and interstitial). FPs were found to be the main primary defects introduced by irradiation at cryogenic temperature also in Ge, and are also stable at least up to 10 K, after this temperature their behavior in n and p-type material being different.

Another type of primary defect, both in silicon and in germanium, is the four-folded coordinated defect (FFCD) [16 – 19]. The vacancy and the interstitial destroy the fourfold coordination of the lattice and relatively high defect formation energy for these defects is the consequence. For both silicon and germanium the defect formation energy is in the order of 3 to 6 eV [17, 20, 21]. The formation energy of FP is less than the sum of an isolated vacancy and interstitial [16]. In contrast to all these point defects, in the FFCD only two bonds are broken, the formation energy is lower in respect to previously mentioned defects, and the bond length and angles do not significantly deviate from their bulk values.

For the case of interest here, of sub-Kelvin conditions, because FFCDs were not clearly confirmed experimentally, we consider that only FPs are formed, and that they do not anneal out in the temperature range where detectors are working. The energy of formation of FPs was calculated in literature using density functional theory both in Si and Ge, as a function of the Fermi level position. Both local density approximation (LDA) and generalized gradient approximation (GGA) were used. The results reported in the literature are summarized in Table 1. For Si, the results are spread in the interval 4.26 – 7.44 eV [16, 22, 23], being dependent on the doping type and on the potential parameters. Less calculations were performed for Ge [24], and lower values were found, 4.2 – 4.9 eV.

Table 1 Energies of formation of FPs in Si and Ge,
calculated using density functional theory

| Formation energy [eV] | Ref. | Obs. |
|---|---|---|
| **Silicon** | | |
| 5.62 | 16 | Close FPs, p-type Si, GGA method |
| 4.32 | 16 | Close FPs, intrinsic Si, GGA |
| 4.26 | 16 | Close FPs, intrinsic Si, LDA |
| 5.77 | 16 | Close FPs, n-type Si, GGA |
| 7.39 | 22 | Distant FPs, intrinsic Si |
| 7.44 | 22 | Distant FPs, n-type Si |
| 6.8 | 23 | Close FPs, distance 4.7 A between constituents, LDA |
| 7.5 | 23 | Close FPs, distance 4.7 A between constituents, GGA |
| **Germanium** | | |
| 4.9 | 24 | Close FPs, distance 4.9 A between constituents, LDA |
| 4.2 | 24 | Close FPs, distance 4.9 A between constituents, GGA-1 |
| 4.2 | 24 | Close FPs, distance 4.9 A between constituents, GGA-2 |

2.2 Contribution of defects to the Luke – Neganov effect

The energy in the electronic system $\eta(E)$ is used in the creation of electron-hole pairs, their number being the ratio between $\eta(E)$ and the energy $\varepsilon$ necessary for the creation of a pair:

$$n(E) = \frac{\eta(E)}{\varepsilon} = \frac{L(E)E}{\varepsilon} \tag{1}$$

where *L(E)* is a factor defined as: $L(E) \equiv \eta(E)/E$, i.e. the fraction of the energy of the recoil transferred to the target as ionization. $\eta(E)$ and $\varepsilon$ respectively could be decomposed into two parts. Reasoning for an electron-hole pair, the first part is related to the production of the pair itself, and corresponds approximately to the semiconductor gap; it is transformed into heat at the electrode during charge collection, by phonon emission. The second is related to the emission of phonons accompanying the production of the electron-hole pair.

The analysis which follows is an extension of the formalism developed in Refs. [25, 26], but which keeps into account also defect formation. In cryogenic detectors, an event is identified using two signatures. The first is the ionization signal, corresponding to the collection at electrodes of the electron-hole pairs created by the energy loss process. The second is the heat (or phonon) signal, recorded by a thermal sensor in contact with the crystalline semiconductor (Ge or Si). The simultaneous measurement of the two signals is an efficient method to discriminate against the background of electron recoils. In the case of electron recoils, all the energy of the recoil is used in the creation of electron-hole pairs, i.e. the corresponding *L(E)* factor equals unity.

Considering the procedure used in Ref. [25], the amplitude of the ionization signal is proportional to the number of electron hole pairs. For electron (gamma) and nuclear recoils, denoted with subscripts '$\gamma$' and '$n$' respectively, this reads as:

$$A_{I,\gamma} \propto n_\gamma = \frac{\eta(E)}{\varepsilon} = \frac{E}{\varepsilon} \tag{2}$$

$$A_{I,n} \propto n_n = \frac{\eta(E)}{\varepsilon} = \frac{L(E)E}{\varepsilon} \tag{3}$$

The amplitude of the signal $A_I$ is usually calibrated using gamma-ray sources to provide $E_I$, the energy in units of keV$_{ee}$, so that:

$$E_{I,n} = L(E)E \tag{4}$$

Considering also the Luke-Neganov effect [27 – 29] in an applied bias $V$, the amplitude of the heat signal is due both to the energy transferred to the lattice, in the form of phonons, by the recoil, and to the energy extracted from the electric field by Joule heating during the drift of collected electrons and holes. For electron recoils:

$$A_{H,\gamma} \propto E + n_\gamma eV = \left(1 + \frac{eV}{\varepsilon}\right)E \tag{5}$$

The amplitude $A_H$ is usually calibrated using gamma-ray sources to provide $E_H$, the energy in units of keV$_{ee}$,

$$E_{H,\gamma} = E. \tag{6}$$

For nuclear recoils, the energy in the electronic system $\eta(E)$ is transformed in heat. Also, the part of the energy transferred to the lattice and which is not stored into defects, $\nu(E) - E_D$, is found as heat. To these two components, one must also add the energy furnished by the applied electric field, so that the amplitude of the heat signal is:

$$A_{H,n} \propto \eta(E) - E_D + \nu(E) + n_n eV = E - E_D + \frac{L(E)E}{\varepsilon}eV = \left(1 + \frac{eV}{\varepsilon}L(E)\right)E - E_D \tag{7}$$

Keeping into account the equations (4) and (5), the heat signal is:

$$E_{H,n} = \frac{1}{1 + \frac{eV}{\varepsilon}}\left[\left(1 + \frac{eV}{\varepsilon}L(E)\right)E - E_D\right] \tag{8}$$

Using the ionization measurement in the form of eq. (4), one obtains for the energy of the recoil:

$$E = E_H\left(1 + \frac{eV}{\varepsilon}\right) - E_I\frac{eV}{\varepsilon} + E_D\left(1 + \frac{eV}{\varepsilon}\right) \tag{9}$$

We would like to underline that $E_D$ depends on $E - E_I$, so that (9) is an implicit formula for *E*. As specified before, this formula is an extension of eq. (6) from Ref. [25], where the energy stored in defects is considered in the energy balance. The energy stored in FPs ($E_D$) is the product between the number of defects, and the energy of formation of a FP. If only the heat signal is measured, then the energy of the recoil is to be obtained by solving for it eq. (8), using, for example, a parameterization for the Lindhard partition factor [9, 30, 31, 32, 33].

## 3. Physical processes related to WIMPs direct detection

The nature and characteristics of DM is a question of central importance in cosmology, astrophysics and astroparticles. The list of candidates and the possible signatures of DM have greatly expanded due to recent experimental results and observations [34, 35]. A summary of dark matter particle candidates, their properties, and the potential methods for their detection was recently given in Refs. [36, 37]. WIMPs are the most studied from all DM candidates, are found in many particle physics theories, have naturally the correct relic density, and could be detected in many ways. For the candidates for WIMPs weak interaction is dominant, they have tree-level interactions with the *W* and *Z* gauge bosons as well as with the gravitational one. No interactions mediated by gluons or photons are permitted. Consequently, they may be directly detected when they scatter off nuclei in terrestrial detectors [38, 39].

WIMPs in dark matter halo move in respect to a terrestrial target with a velocity in the range 230 – 260 km/s. Its motion is composed from the galactic motion, the Sun mean motion relative to nearby stars and the Earth's orbital motion relative to the Sun [40]. The value of 260 km/s will be used in this paper as the average velocity of DM particles in respect to the detector.

In the SUSY models, masses for WIMPs are in the range around of the Weak scale (100 GeV), but light neutralinos with masses in the keV to GeV range [41] remain an interesting possibility, theoretically motivated if DM does not couple strongly to the visible sector. Many existing models can accommodate light DM; see for example Refs. [41 – 44] and references cited therein. In this paper, we consider WIMP masses in the range 5 – 100 GeV.

For nonrelativistic WIMPs particles with an arbitrary spin, in the approximation of Fermi's Golden Rule, the formula for the cross section contains spin-independent (mostly scalar) and spin-dependent (mostly axial vector) terms [35]. Details for coherence /decoherence conditions for cross sections as well as the details for spin independence/dependence of the cross sections are discussed in details in the cited paper.

## 4. Results and discussion

In the discussion which follows, a WIMP with mass in the range 5 -100 GeV, having a velocity of 260 km/s in respect to a terrestrial detector is considered. It has a single interaction in a Ge or Si cryogenic detector.

For silicon, the first result of the energy partition between ionization and other processes using the complete Lindhard theory was obtained by Lindhard and later published in the paper of Simon [45]. Analytical approximations of the Lindhard equations both for silicon and germanium are reported by Lazanu and Lazanu [31]. Robinson [30], starting from Lindhard's asymptotic equations, gave the parameters of the partition factor between ionization and atomic collisions. Akkerman and Barak [46] claim that previous calculations overestimate the electronic losses at low energies, below 100 keV, and provide a correction to the partition factor in silicon, which consists in new parameters for the same form of the partition factor as the one suggested by Lindhard and used by Robinson.

The results for the dependence of the energy imparted to the atomic system of Si as a function of recoil's energy are represented in Figure 1. In the range of energies of interest, of tens of keV, a good agreement between all calculations reported in the literature can be seen.

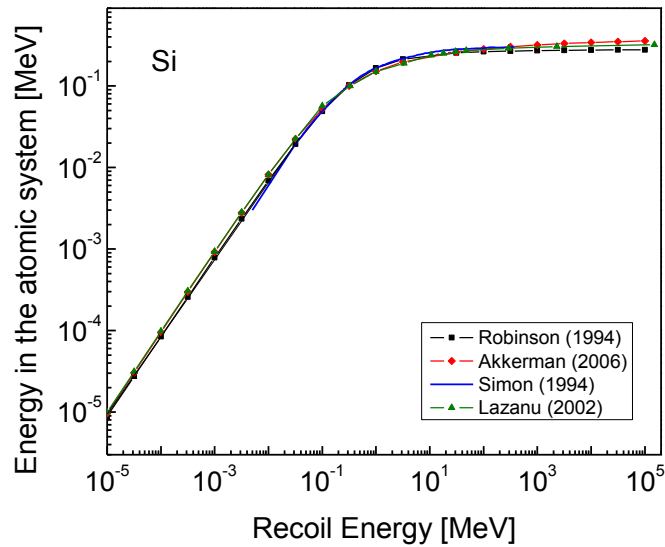

Fig. 1: Lindhard curves for Si

For germanium, in the calculation of the partition factor, we used Robinson's [30] formula, together with the results reported by Lazanu and Lazanu [31], and the parameters provided by Akkerman and Barack [46] for Si.

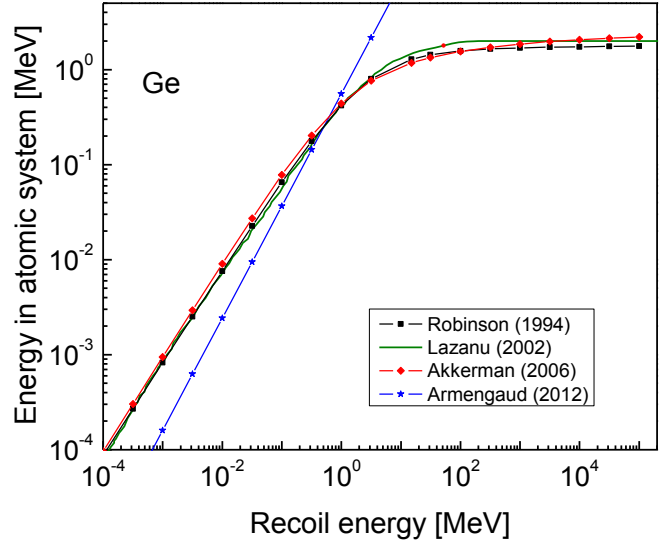

Fig. 2: Lindhard curves for Ge

Armengaud *et al*. [33] uses a power law dependence for the nuclear recoil ionization fraction versus recoil energy, using the idea of quenching factor, result that does not agree with Lindhard's theory in this case. In Fig. 2 we present comparatively all the results for the partition factor for germanium. Scopel [47] underlines the discrepancies between this method of determination of energy partition and Lindhard theory results for germanium. A very useful review of experimental data for the ionization efficiency (calculated from the quenching factor) is presented in Ref. [48], together with a comparison of the data with the results of different proposed models, including also Lindhard's theory for different values of the *k* factor (see Ref. [9] for the definition of this factor). The data points measured in different experiments do not entirely agree, and are affected by large systematic errors [48, 49, 50]. More, although the experimental data points are in a reasonable agreement with the models, the general dependencies of experimental data and models as a function of the recoil energy are different - see Fig. 7 of Ref. [48].

The discrepancies between the measured data and theory, pointed out also for LXe [51] and Si [52] could be due to more aspects which must be further analyzed: a) For recoil energies lower than tens of keV the accuracy of the calculated partition factor decreases, due to the fact that probably both the nuclear and the electronic stopping powers are not adequately described in Lindhard's theory for slow recoils. In what regards the nuclear stopping power, the problem is related mainly to the screened interatomic potential [53]. The difficulties in the modeling of the electronic stopping power for slow recoils were reviewed by Sigmund in Ref. [54]. b) The state of ionization of the recoil (effective charge) is dependent on the relation between its velocity and the Bohr velocity of the electrons. c) All calculations in the frame of Lindhard's theory are based on the hypothesis of amorphous targets. In crystals, the threshold energy for displacements has different values, as a function of the direction in respect to the lattice orientation.

The average number of displacements (vacancy – interstitial pairs) produced by a recoil of energy $E$ could be estimated based on the Kinchin and Pease damage function [55], which is directly related to the energy of the recoil. The modified Kinchin-Pease function [56, 57] is based on the energy imparted to the atomic system, calculated in its turn using Lindhard's partition factor. The results are presented in Figures 3 and 4 for Si and Ge respectively. The double axis permits a simultaneous reading of the number of defects produced and of the energy deposited in these defects. The lowest values for the formation energy of FPs were considered, in order to calculate a minimum value for the energy stored in these defects. For Si, we utilised the value 4.29 eV in intrinsic material, as the average between the LCA and GGA calculations. For Ge, the value of 4.5 eV was used (see Table 1). In these estimations, as a first approximation, we neglect replacement collisions, when distant Frenkel pairs could be produced (for this case the formation energy in intrinsic silicon being 7.39 eV – see Table 1).

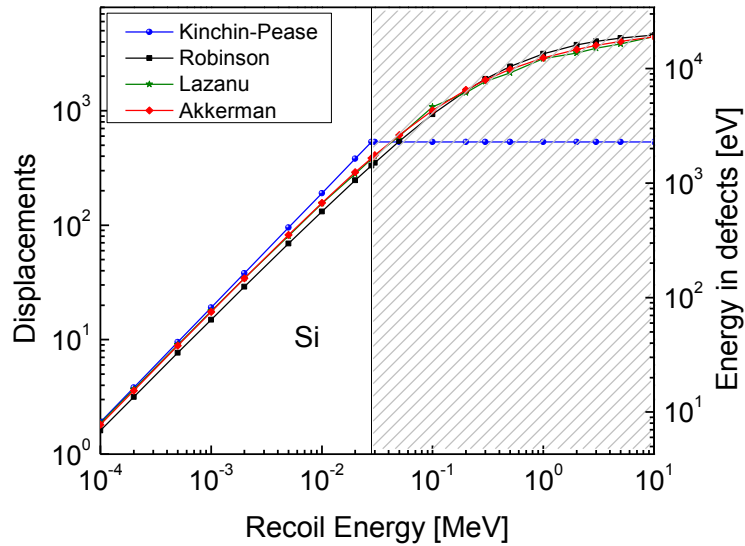

Fig. 3: Number of FPs (left) and energy stored in them (right) produced by a selfrecoil in Si, versus recoil's energy

In Figures 3 and 4, the energies allowed for the recoil produced by a WIMP with the characteristics specified before are evidenced, while the area corresponding to higher recoil energies are hatched.

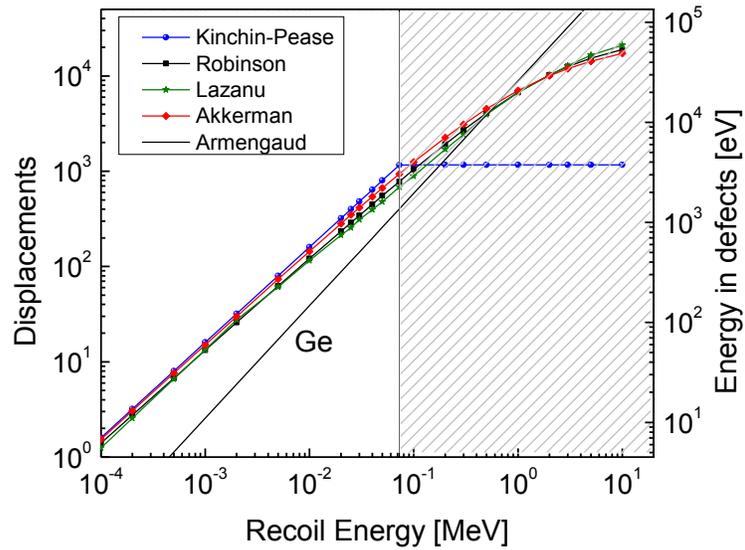

Fig. 4: Number of FPs (left) and energy stored in them (right) produced by a selfrecoil in Ge, versus recoil's energy

Figures 5 and 6 illustrate the dependence of the energy stored in FPs as a function of the mass of the WIMP and the centre of mass (CM) scattering angle, for a hypothetical WIMP moving in respect to the detector with a velocity of 260 km/s, and which interacts only once in the detector. Both for Si and Ge, Robinson's (1994) partition factor was used.

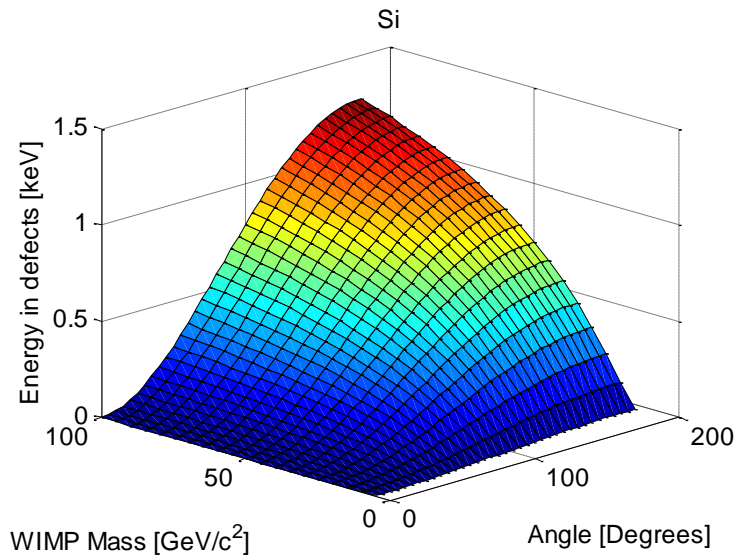

Fig. 5: Dependence of the energy stored in the defects produced in Si by a WIMP interaction as a function of WIMP's mass and of the CM scattering angle

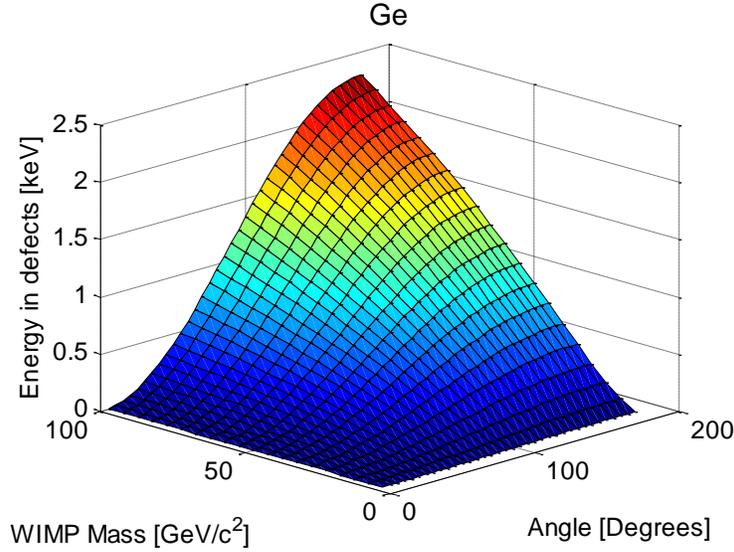

Fig. 6: Dependence of the energy stored in the defects produced in Ge by a WIMP interaction as a function of WIMP's mass and of the CM scattering angle

The maximum energy stored in the defects produced by the interaction of a WIMP in the mass range considered with Si and Ge at cryogenic temperatures is around 1.27 and 2.3 keV respectively.

In accordance with Luke – Neganov prescriptions, the energy stored in defects, in an applied bias has a contribution multiplied by the factor $\left(1 + \dfrac{eV}{\varepsilon}\right)$.

For practical situations, particles from the radioactive background (muons, neutrons, etc.) or from products of reactions induced in the material of detectors could represent supplementary sources of defects. The partition of the energy of a selfrecoil, discussed in this work, must then be supplemented by the partition of energies of those particles (with lower mass and charge number than the target) [58, 59], evaluating this way a supplementary energy stored in defects.

**Summary**

The possibility of defect formation in bolometric semiconductor detectors at cryogenic temperatures was studied, with application to WIMPs direct searches.

The models for the partition factor between the energy transferred by the primary recoil to the atomic and electronic systems of Si and Ge were reviewed, starting from Lindhard's theory. Part of the energy transferred to the atomic system is stored in defects. At sub-Kelvin temperatures the defects are Frenkel pairs and they do not anneal out.

Considering the energy stored in defects, the Luke–Neganov formula was extended. The energy deposited in stable defects by the recoils produced in the elastic interactions of WIMPs with target atoms in both silicon and germanium was estimated. Its value is up to 1.27 keV for silicon and up to about 2.3 keV for germanium respectively, for WIMPs of mass 5-100 GeV/$c^2$, moving with a velocity of 260 km/s in respect to the detector.

## Acknowledgements

MLC and SL thank the NIMP Core Programme PN09-450101 for financial support.

References

1. F. Simon, Application of Low Temperature Calorimetry to Radioactive Measurements, Nature 135, (1935) 763.
2. R.J. Gaitskell, Direct detection of dark matter, Annu. Rev. Nucl. Part. Sci. 54 (2004) 315–359.
3. X. Sarazin, Review of double beta experiments [arXiv: 1210.7666].
4. G. J. Yong., R. M. Kolagani, S. Adhikari, R. M. Mundle, D. W. Cox, A. L. Davidson III, Colossal Magnetoresistive Manganite Based Fast Bolometric X-ray Sensors for Total Energy Measurements of Free Electron Lasers, LLNL-JRNL-409505 (2008).
5. S. Marnieros, L. Dumoulin, A. Benoit, L. Berge, P. Camus, S. Collin, A. Juillard.,C.A. Marrache-Kikuchi, All electron bolometer for radiation detection, J. Phys.: Conf. Ser. 150 (2009) 012027.
6. K. Karsten K. and P. Ehrhart, Frenkel pairs in low-temperature electron-irradiated InP: X-ray diffraction, Phys. Rev. B 51 (1995) 10508-10519.
7. H. Zillgen, P. Ehrhart, Bound vacancy interstitial pairs in irradiated silicon, Nucl. Instr. Meth. Phys. Res. B 127/128 (1997) 27-31.
8. V. V. Emtsev V. V., A. M. Ivanov, V. V. Kozlovski., A. A. Lebedev, G. A. Oganesyan, N. B. Strokan, and G. Wagner, Similarities and Distinctions of Defect Production by Fast Electron and Proton Irradiation: Moderately Doped Silicon and Silicon Carbide of n-Type, Semiconductors 46 (2012) 456–465.
9. J. Lindhard, V. Nielsen, M. Scharff, P.V. Thomsen, Integral equations governing radiation effects, Mat. Phys. Medd. Dan Vid. Sesk. 33 (1963) 1-42.
10. P M Fahey, P B Griffin and J D Plummer, Point defects and dopant diffusion in silicon, Rev. Mod. Phys. 61 (1989) 289-384.
11. P. Ehrhart P. and H. Zillgen, Vacancies and interstitial atoms in e⁻-irradiated germanium, J. Appl. Phys. 85 (1999) 3503-3511.
12. A. Polity, F. Rudolf, Defects in electron-irradiated Ge studied by positron lifetime spectroscopy, Phys Rev B 59 (1999) 10025-10030.
13. P. Ehrhart P. and H. Zillgen, Vacancies and interstitial atoms in irradiated silicon, Mat. Res. Soc. Symp. Proc. 469 (1997) 175-177.
14. V. V. Emtsev, P. Ehrhart, D. S., Poloskin, K. V. Emtsev, Comparative studies of defect production in heavily doped silicon under fast electron irradiation at different temperatures, J Mater Sci: Mater Electron 18 (2007) 711–714.
15. V. Emtsev, Point defects in germanium: Reliable and questionable data in radiation experiments, Mat. Sci. in Sem. Proc. 9 (2006) 580–588.

16. S. Goedecker S, Th. Deutsch and L. Billard, A Fourfold Coordinated Point Defect in Silicon. Phys. Rev. Lett. 88 (2002) 235501 [4 pages].

17. I. Lazanu, S. Lazanu, The role of primary point defects in the degradation of silicon detectors due to hadron and lepton irradiation, Phys. Scr. 74 (2006) 201-207.

18. M. D. Moreira, R. H. Miwa, P. Venezuela, Electronic and structural properties of germanium self-interstitials Phys. Rev. B 70 (2004) 115215 [5 pages].

19. D. Caliste, P. Pochet, T. Deutsch and F. Lançon, Germanium diffusion mechanisms in silicon from first principles, Phys. Rev. B 75 (2007) 125203 [6 pages].

20. N. Fukata, A. Kasuya and M. Suezawa, Formation energy of vacancy in silicon determined by a new quenching method, Physica B 308-310 (2002) 1125-1126.

21. P. Spiewak K. J. Kurzydowski, J. Vanhellemont, P. Clauws, P. Wabinski, K. Mynarczyk, I. Romandic, and A. Theuwis, Simulation of intrinsic point defect properties and vacancy clustering during Czochralski germanium crystal growth, Mat. Sci. Sem. Proc. 9 (2006) 465-470.

22. S. A. Centoni., B. Sadigh., G.H. Gilmer, T.J. Lenosky, T. Díaz de la Rubia, and C.B. Musgrave., First-principles calculation of intrinsic defect formation volumes in silicon. Phys. Rev B 72 (2005) 195206 [9 pages].

23. E. Holmström E., A. Kuronen, and K. Nordlund, Threshold defect production in silicon determined by density functional theory molecular dynamics simulations Phys. Rev. B 78 (2008) 045202 [6 pages].

24. E.Holmström, K. Nordlund and A. Kuronen, Threshold defect production in germanium determined by density functional theory molecular dynamics simulations, Phys. Scr. 81 (2010) 035601[4 Pages]

25. A. Benoit *et al.*, EDELWEIS Collab., Measurement of the response of heat-and-ionization germanium detectors to nuclear recoils, Nucl. Instr. Meth. Phys. Res. A 577 (2007) 558‑568.

26. V. Sanglard *et al.*, EDELWEISS Collab., Final results of the EDELWEISS-I dark matter search with cryogenic heat-and-ionization Ge detectors, Phys. Rev. D 71 (2005) 122002 [16 pages].

27. B. Neganov and V. Trofimov, USSR patent No 1037771, 1981.

28. P. N. Luke, Voltage-assisted calorimetric ionization detector. J. Appl. Phys. 64 (1988) 6858-6860.

29. M. P. Chapellier, G. Chardin, L. Miramonti, X. Francois Navick, Physical interpretation of the Neganov-Luke and related effects, Physica B 284-288 (2000) 2135-2136.

30. M. T. Robinson, Basic physics of radiation damage production, J. Nucl. Mat. 216 (1994) 1-28.

31. S. Lazanu, I. Lazanu, Analytical approximations of the Lindhard equations describing radiation effects, Nucl. Instr. Meth. Phys. Res. A 462 (2001) 530–535.

32. Z. Ahmed *et al.*, CDMS Collab., Results from a Low-Energy Analysis of the CDMS II Germanium Data, Phys. Rev. Lett. 106 (2011) 131302 [5 pages] [arXiv:1011.2482].

33. E. Armengaud *et al*., EDELWEISS Collab., A search for low-mass WIMPs with EDELWEISS-II heat-and-ionization detectors, Phys. Rev. D 86 (2012) 051701(R) [arXiv:1207.1815].

34. E. Komatsu E., K. M.Smith, J.Dunkley, C. L. Bennett, B. Gold., G. Hinshaw, N. Jarosik, D. Larson, M. R. Nolta, L. Page, D. N. Spergel., M. Halpern, R. S. Hill, A. Kogut, M. Limon, S.S. Meyer, N. Odegard, G. S. Tucker, J. L. Weiland, E. Wollack, and E. L. Wright, Seven-Year Wilkinson Microwave Anisotropy Probe (WMAP) Observations: Cosmological Interpretation, ApJ Suppl.192 (2011) 18 [47 pages] [arXiv:1001.4538].

35. N.Jarosik, C. L. Bennett, J.Dunkley, B. Gold, M.R. Greason., M. Halpern, R. S. Hill, G. Hinshaw, A. Kogut, E. Komatsu, D. Larson, M. Limon, S. S. Meyer, M. R. Nolta, N. Odegard, L. Page, K. M. Smith, D. N. Spergel, G. S. Tucker, J. L. Weiland, E. Wollack, E. L. Wright, Seven-Year Wilkinson Microwave Anisotropy Probe (WMAP) Observations: Sky Maps, Systematic Errors, and Basic Results, ApJ. Suppl. 192 (2011) 14-[15 pages] [arXiv:1001.4744].

36. J. L. Feng, Dark Matter Candidates from Particle Physics and Methods of Detection, Ann. Rev. Astron. Astrophys. 48 (2010) 495–545.

37. R. W. Schnee, Introduction to Dark Matter Experiments, Proc. 2009 Theoretical Advanced Study Institute in Elementary Particle Physics, (World Scientific, Singapore), pp. 629–681, [arXiv:1101.5205v1].

38. M. W. Goodman and E. Witten, Detectability of certain dark matter candidates, Phys. Rev. D 31 (1985) 3059-3063.

39. J R. Primack, D. Seckel, and B. Sadoulet, Detection of Cosmic Dark Matter, Ann. Rev. Nucl. Part. Sci. 38 (1988) 751-807.

40. M.T. Frandsen, F.Kahlhoefer, C. McCabe, S. Sarkara and K. Schmidt-Hoberg, Resolving astrophysical uncertainties in dark matter direct detection. JCAP 01 (2012) 024 [30 pages] [arXiv:1111.0292].

41. R. Essig, J. Mardon, T. Volansky, Direct Detection of Sub-GeV Dark Matter, Phys.Rev. D85 (2012) 076007 [9 pages][arXiv:1108.5383].

42. H. T. Wong, Ultra-Low-Energy Germanium Detector for Neutrino-Nucleus Coherent Scattering and Dark Matter Searches, Mod.Phys.Lett.A23 (2008)1431-1442 [arXiv:0803.0033].

43. B. Dumont, G. Belanger, S. Fichet, S. Kraml, T. Schwetz, Mixed neutrino dark matter in light of the 2011 XENON and LHC results, JCAP (2012) [arXiv:1206.1521].

44. Y. S. Jeong, C. S. Kim., M. H. Reno, Majorana Dark Matter Cross Sections with Nucleons at High Energies [arXiv:1207.1526].

45. G. W. Simon, J. M. Denney, R. G. Downing, Energy Dependence of Proton Damage in Silicon, Phys. Rev. 129 (1963) 2454-2459.

46. A. Akkerman A. and J. Barak, New Partition Factor Calculations for Evaluating the Damage of Low Energy Ions in Silicon, IEEE Trans Nucl. Sci. 53 (2006) 3667-3674.

47. S. Scopel, The WIMP Interpretation of DAMA, CoGeNT and CRESST: The Dark at the End of the Tunnel?, talk at KIAS Phenomenology Workshop, 17-19 November 2011.

48. D. Barker and D.-M. Mei, Germanium Detector Response to Nuclear Recoils in Searching for Dark Matter, [arXiv:1203.4620].

49. P S Barbeau, J. I. Collar and O Tench, Large-mass ultralow noise germanium detectors: performance and applications in neutrino and astroparticle physics, JCAP 09 (2007) 2009.

50. C.E. Aalseth *et al.*, CoGeNT Collab., A Search for Low-Mass Dark Matter using p-type Point Contact Germanium Detectors [arXiv:1208.5737].

51. P. Sorensen, C.E. Dahl, Nuclear recoil energy scale in liquid xenon with application to the direct detection of dark matter, PRD 83 (2011) 063501 [6 pages].

52. J. Barreto H. Cease, H.T. Diehl, J. Estrada, B. Flaugher, N. Harrison J. Jones, B. Kilminster, J. Molina, J. Smith, T. Schwarz and A. Sonnenschein, Direct Search for Low Mass Dark Matter Particles with CCDs, Phys. Lett. B 711, (2012) 264-269 [arXiv:1105.5191].

53. M. Nastasi, J.W. Mayer, J.K. Hirvonen, Ion-solid interactions: fundamentals and applications, Cambridge University Press 1996.

54. P. Sigmund, Stopping of slow ions, Bulletin of the Russian Academy of Sciences: Physics, 72 (2008) 569–578.

55. G. H. Kinchin, R. S. Pease, The displacement of atoms in solids by radiation. Rep. Progr. Phys. 18 (1955) 1-51.

56. P. Sigmund, On the number of atoms displaced by implanted ions or energetic recoil atoms, Appl. Phys. Lett. 14 (1969) 114-117.

57. M. T. Robinson and O. S. Oen, On the use of thresholds in damage energy calculations, J. Nucl. Mater. 110 (1982) 147-149.

58. A. Van Ginneken, Non Ionizing Energy Deposition in Silicon for Radiation Damage Studies, Fermi National Accelerator Laboratory FN-522 (1989).

59. S. Lazanu, I. Lazanu, Si, GaAs and diamond damage in pion fields with application to LHC, Nucl. Instr. Meth. Phys. Res. A 419 (1998) 570-576.